# Ferromagnetic Kondo lattice behavior in $Ce_{11}Pd_4In_9$

Debarchan Das and Dariusz Kaczorowski*

Institute of Low Temperature and Structure Research, Polish Academy of Sciences,

P.O.Box 1410, 50-950 Wrocław, Poland

**Abstract**

We report on the low-temperature physical properties of a novel compound $Ce_{11}Pd_4In_9$ that crystallizes with the orthorhombic $Nd_{11}Pd_4In_9$-type crystal structure (space group *Cmmm*). The compound exhibits ferromagnetic ordering at $T_C$ = 18.6 K and an order-order transition at $T_t \approx$ 1.6 K, as inferred from the low-temperature magnetic susceptibility, heat capacity and electrical resistivity data. In the paramagnetic region, the electrical transport in $Ce_{11}Pd_4In_9$ is dominated by Kondo effect. Below $T_C$, a distinct contribution due to ferromagnetic spin waves dominates the electrical resistivity data, while at the lowest temperatures, the electrical transport and thermodynamic properties are governed by strong electron-electron correlations. The features observed conjointly hint at strongly correlated ground state in $Ce_{11}Pd_4In_9$.

* Corresponding author.

E-mail address: D.Kaczorowski@int.pan.wroc.pl (Dariusz Kaczorowski)

## 1. Introduction

For past few decades, investigation of cerium-based intermetallic compounds has been at the forefront of condensed matter research as it encompasses a large variety of interesting phenomena which include Kondo effect, heavy-fermion (HF) behavior, non-Fermi liquid (NFL) features, magnetic ordering, superconductivity, quantum criticality, etc. [1-5]. Hybridization between 4$f$ localized electrons and conduction electrons plays a crucial role in governing the ground state properties in such materials. Of particular interest are Kondo lattice systems which exhibit magnetic ordering, as in many of them one can tune the magnetic exchange interactions by external parameter, like pressure, magnetic field or doping, to explore the possibility of suppressing the magnetic order down to absolute zero temperature at a quantum critical point (QCP). Since a few years, quantum criticality in HF systems has been one of the hot topics in condensed matter physics, owing to unusual phenomena observed in the vicinity of QCP, such as unconventional superconductivity and NFL behavior [2,5,6]. Unlike antiferromagnetic (AFM) dense Kondo systems, examples of ferromagnetic (FM) Kondo lattices are quite rare in the existing literature. Representatives of the latter group, discovered in recent years, are the FM compounds CeRuPO [7], $Ce_3RhSi_3$ [8], $CePd_2Al_8$ [9] and $CeCrGe_3$ [10,11].

Another subject that recently is attracting much research interest concerns Ce-based Kondo lattices that host more than one inequivalent Ce site in their crystallographic unit cell. This structural feature may bring about complex physical properties, like separate AFM order in two Ce atom sublattices in $Ce_5Ni_6In_{11}$ [12], dipolar and quadrupolar AFM orderings in $Ce_3Pd_{20}Si_6$, each associated with different Kondo ion site [13, 14], or coexistence of AFM and HF superconductivity in $Ce_3PtIn_{11}$, where the two cooperative phenomena likely emerge in two distinct Ce atom sublattices [15].

Within the two frameworks outlined above, search for novel Ce-based intermetallics with multiple Ce sites in their crystal structure, which exhibit Kondo effect and possibly order ferromagnetically at low temperatures appears to be of particular interest. Recently, a series of rare-earth (RE) ternaries with the chemical formula $RE_{11}T_4In_9$ (T = $d$-electron transition metal) has been reported to form with orthorhombic crystal structure of the $Nd_{11}Pd_4In_9$-type (space group $Cmmm$) that hosts as many as five independent RE sites in the unit cell [16-19]. Amidst these compounds, two Ce-based materials were found. Most interestingly, $Ce_{11}Ni_4In_9$ was established to order ferromagnetically below $T_C$ = 16.5 K and show an order-order phase transition at $T_t$ = 5 K [20], while $Ce_{11}Ru_4In_9$ was characterized as a ferrimagnet with $T_C$ = 6.3

K and weak Kondo effect [21]. Motivated by these findings, we successfully synthesized another compound from this family, namely $Ce_{11}Pd_4In_9$, the existence of which was briefly communicated in Ref. 18. In this paper, we report on the low temperature physical properties of the novel phase, studied by means of magnetic, electrical resistivity and heat capacity measurements. Remarkably, the compound was found to be a moderate HF system that orders ferromagnetically below $T_C$ = 18.6 K and exhibits another magnetic phase transition in the ordered state.

## 2. Experimental

Polycrystalline sample of $Ce_{11}Pd_4In_9$ was synthesized by arc-melting elemental constituents (purities: Ce-3N, Pd-4N, and In-6N) on a water-cooled copper hearth in an arc furnace installed inside a glove-box filled with ultra-pure argon gas with continuously controlled partial pressures of $O_2$ and $H_2O$ to be lower than 1 ppm. The ingot was flipped over and remelted several times to ensure homogeneity. The weight loss after the final melting was negligible (less than 0.4%). Annealing at $700^0$C for two weeks resulted in producing microcracks in the sample that preserved the same crystal structure. Thus, we restricted our measurements only to the as-cast material. As a nonmagnetic reference compound, $La_{11}Pd_4In_9$ (La metal purity: 3N) was also synthesized following the same process.

Quality of the obtained alloys was checked by x-ray powder diffraction (XRD) at room temperature on an X'pert Pro PanAnalytical diffractometer using Cu-$K_\alpha$ radiation. The XRD data confirmed that $Ce_{11}Pd_4In_9$ and $La_{11}Pd_4In_9$ crystallize with the orthorhombic $Nd_{11}Pd_4In_9$-type crystal structure, and the obtained lattice parameters were in agreement with the literature data [18]. Chemical composition of the samples was examined by energy dispersive x-ray (EDX) analysis using a FEI scanning electron microscope equipped with an EDAX Genesis XM4 spectrometer. The results indicated that the obtained polycrystalline samples are homogeneous and single-phase materials with the stoichiometry close to the nominal one.

Magnetic measurements were performed in the temperature range 1.72 – 300 K using a Quantum Design SQUID magnetometer. The electrical resistivity was measured over the temperature range 0.38 – 300 K employing standard ac four-probe technique implemented in a Quantum Design PPMS platform. Specific heat measurements were carried out from 0.4 to 300 K using relaxation method available in the same PPMS platform.

## 3. Results

*3.1. Magnetic properties*

The results of magnetic measurements performed on $Ce_{11}Pd_4In_9$ are summarized in Fig. 1. Above about 150 K, the reciprocal magnetic susceptibility follows the Curie-Weiss law with the effective magnetic moment $\mu_{eff}$ = 2.62(5) $\mu_B$ per Ce atom and the paramagnetic Curie temperature $\theta_p$ = –9.0(2) K. The experimental value of $\mu_{eff}$ is close to the theoretical prediction for free trivalent cerium ion (2.58 $\mu_B$). Its magnitude, calculated by averaging the separate contributions from five independent magnetic sublattices, suggests similar valence state of each Ce ion in the unit cell of $Ce_{11}Pd_4In_9$. In turn, the fairly large negative value of $\theta_p$ hints at predominance of antiferromagnetic correlations in the compound that orders ferromagnetically (see below). It seems likely that $\theta_p$ reflects significant Kondo interactions resulting in antiferromagnetic correlations in the system, alike in ferromagnetic $CePd_2Al_8$ [9], $CeIr_2B_2$ [22] and $CeTiGe_3$ [23]. Below 150 K, the $\chi^{-1}(T)$ dependence markedly deviates from a straight-line behavior evidencing strong crystalline electric field (CEF) effect.

As shown in the inset to Fig. 1, the low-temperature dependence of the magnetization in $Ce_{11}Pd_4In_9$, measured in an external magnetic field of 1 kOe (upon cooling the specimen in the applied field) exhibits a behavior characteristic of ferromagnets. The Curie temperature, defined as an inflection point on the $M(T)$ curve, equals $T_C$ = 18.6(4) K. With decreasing temperature in the ordered state, the magnetization saturates at a value of 13.7 emu/g that corresponds to the magnetic moment of 0.67(3) $\mu_B$ per Ce ion, calculated assuming equal contribution from each Ce site.

The magnetization measured at $T$ = 1.72 K as a function of applied magnetic field strength (see Fig. 2) corroborates the ferromagnetic nature of the electronic ground state in $Ce_{11}Pd_4In_9$. Remarkably, in weak magnetic fields, the magnetization rapidly increases with rising field reaching about 13 emu/g in $H$ = 1 kOe, and then almost saturates at this value up to 4 kOe (the inset to Fig. 2). In stronger fields, the magnetization increases gradually with a clear tendency for saturation. In the limiting field of 70 kOe, it achieves a value of about 21 emu/g yielding the average magnetic moment of 1.02(3) $\mu_B$ per Ce site. This value is only a fraction of that expected for a free trivalent Ce ion ($gJ$ = 2.15 $\mu_B$), and must be attributed to the doublet ground state in the CEF split $^2F_{5/2}$ multiplet. In parallel, some reduction of the ordered magnetic moment due to the Kondo interactions can be also expected (see below). As can be inferred from the inset to Fig. 2, the characteristic feature of ferromagnetic $Ce_{11}Pd_4In_9$ is a large remanence (~60%) combined with a fairly narrow hysteresis loop, roughly limited to magnetic fields $H < 1$ kOe at $T = 1.72$ K.

In order to gain better understanding of the magnetic state in $Ce_{11}Pd_4In_9$, magnetization measurements were performed in a weak magnetic field of 20 Oe, upon cooling the sample in zero (ZFC: zero field cooling) and applied (FC: field cooling) field. As can be inferred from Fig. 3, the obtained $M(T)$ data bifurcates at a temperature slightly lower than $T_C$, as expected for ferromagnets. However, worth noting is an unusual shape of the FC curve, which deviates at the bifurcation point from a Brillouin function and becomes hardly temperature dependent. Similarly, the ZFC curve, which can be expected to be governed by domain effects (hence the negative sign of the signal measured in this regime), does not behave in a manner known for simple ferromagnets, and exhibits a pronounced hump near 7 K. Both findings hint at a more complex FM ordering in $Ce_{11}Pd_4In_9$ than a simple parallel arrangement of all the cerium magnetic moments present in the unit cell. It seems likely that the complexity of the ordered state in the compound studied is related to the occurrence of as many as five distinct magnetic sublattices, which may slightly differ in their individual contributions to the macroscopic magnetization observed below $T_C$. Alternatively, however, the abnormal $M(T)$ variations in weak magnetic fields may originate from a specific interplay between the magnetic exchange, magnetocrystalline and magnetic energies. To clarify the actual nature of these anomalies, angle-dependent magnetization measurements on single crystals of $Ce_{11}Pd_4In_9$ are compulsory.

The magnetic susceptibility measurements carried out for the compound $La_{11}Pd_4In_9$, which was used in this study as a reference material to $Ce_{11}Pd_4In_9$ (without 4$f$ electrons), revealed that it is a weak Pauli paramagnet with the molar magnetic susceptibility of the order of $10^{-5}$ emu/mol per La atom at room temperature.

## 3.2. Heat capacity

To elucidate the low-temperature thermodynamic properties of $Ce_{11}Pd_4In_9$, the heat capacity measurements were performed on the very same sample of this material that was studied for its magnetic behavior (see above). Furthermore, the non-magnetic counterpart $La_{11}Pd_4In_9$ was measured, and the obtained data are shown in Fig. 4. As expected, the specific heat of the latter compound varies with temperature in a manner typical for metallic paramagnets. At room temperature, it is close to the Dulong-Petit limit $3nR = 598.6$ J/(mol K) [where $n$ represents the number of atoms per formula units (here $n = 24$), and $R = 8.314$ J/(mol K)] is the universal gas constant]. In the entire temperature range studied, the $C(T)$ curve can be approximated by the function [24]

$$C(T) = \gamma T + 9nR\left(\frac{T}{\Theta_D}\right)^3 \int_0^{\frac{\Theta_D}{T}} \frac{x^4 e^x dx}{(e^x-1)^2} \quad , \tag{1}$$

where the first term represents the contribution due to conduction electrons ($\gamma$ is the Sommerfeld coefficient which was fixed at 56.4 mJ/mol K$^2$; see below), and the second term stands for the lattice contribution described within the Debye model ($\Theta_D$ is the Debye temperature). Least-squares fitting Eq. 1 to the experimental data of La$_{11}$Pd$_4$In$_9$ yielded $\Theta_D$ = 172.8(9) K.

The low-temperature ($T < 3$ K) specific heat data of La$_{11}$Pd$_4$In$_9$ were analyzed also in terms of the standard formula $C(T) = \gamma T + \beta T^3$, where $\beta = 12\pi^4 nR/5\theta_D^3$, and the result is displayed in the inset to Fig. 4. The so-obtained parameters are $\gamma = 56.4(7)$ mJ/(mol K$^2$) and $\beta = 4.2809(3)$ mJ/(mol K$^4$), yielding $\Theta_D$ = 221(9) K. It is worth noting that $\gamma$ is close to that reported before for La$_{11}$Ni$_4$In$_9$ [74.8mJ/(mol K$^2$), Ref. 19]. In turn, the difference between the two estimates for the Debye temperature can be considered reasonable in view of the crudeness of the two fitting approaches applied. As can be inferred from the inset to Fig. 4, at the lowest temperatures studied ($T < 0.5$ K), the $C/T$ ratio of La$_{11}$Pd$_4$In$_9$ deviates from the $T^2$ behavior, probably signaling the contribution due to nuclear heat capacity.

Fig. 5 shows the temperature dependence of the specific heat of Ce$_{11}$Pd$_4$In$_9$. As emphasized in the inset to this figure, there occurs a prominent λ-type anomaly caused by the ferromagnetic phase transition at $T_C$ = 18.6 K. However, jump of the heat capacity at $T_C$ is found out to be only about 5.6 J/(mol K) per Ce atom, which is much less than $\Delta C = 5RS(S+1)/[S^2+(S+1)^2] = 12.47$ J/(mol K), calculated within the molecular field approximation (MFA) for a simple ferromagnet with the effective spin $S$ = ½. The reduction probably arises due to Kondo screening interactions. Considering the $S$ = ½ resonant model [25-27], $\Delta C$ can be related to the characteristic Kondo temperature $T_K$ via the equation

$$\Delta C = \frac{6k_B}{\psi'''\left(\frac{1}{2}+\zeta\right)}\left[\psi'\left(\frac{1}{2}+\zeta\right) + \zeta\psi''\left(\frac{1}{2}+\zeta\right)\right]^2 \tag{2}$$

where $\zeta = (T_K/T_C)/2\pi$ and $\psi'$, $\psi''$ and $\psi'''$ are the first three derivatives of the digamma function, respectively. This universal relationship yields for $\Delta C$ = 5.6 J/(mol$_{Ce}$ K), observed for Ce$_{11}$Pd$_4$In$_9$, the ratio $T_K/T_C$ = 1.03, and thus $T_K$ of about 19 K.

The remarkable feature in the heat capacity data of Ce$_{11}$Pd$_4$In$_9$ is the occurrence of another pronounced singularity in the ordered region in the form of a sharp nearly-symmetric feature in $C/T(T)$, centered at $T_t$ = 1.6 K. This anomaly manifests the first-order type transition, likely a change in the magnetic structure, e.g. from a noncollinear FM (see the remark above) to the collinear FM one. Neutron diffraction experiments are called for to unveil the actual arrangements of the magnetic moments in $Ce_{11}Pd_4In_9$ above and below $T_t$.

The *4f*-electron contribution to the heat capacity of $Ce_{11}Pd_4In_9$ was calculated by subtracting from the measured $C(T)$ data the phonon contribution $C_{ph}(T)$ shown in Fig. 5 by the solid line. The latter variation was assumed to be similar to the phonon term obtained for the nonmagnetic reference material $La_{11}Pd_4In_9$ (the proper mass correction was made in the standard manner). The so-estimated dependence $C_{4f}(T)$ is presented in Fig. 6. A broad hump around 150 K can be associated with CEF splitting of the cerium $^2F_{5/2}$ ground multiplet. However, as each of the five inequivalent Ce atom sites in the orthorhombic unit cell of $Ce_{11}Pd_4In_9$ may have a different scheme of the CEF energy levels, analysis of this anomaly using a simple Schottky model with three Kramers doublets is not applicable. Due to the occurrence of the peak at 1.6 K, the correct estimation of Sommerfeld coefficient value is not possible. However, considering the ratio $C_{4f}/T = 194$ mJ/(mol$_{Ce}$ K$^2$) at lowest temperature investigated ($T = 0.38$ K), one cannot rule out the possibility of formation of a HF ground state in the compound investigated.

Shown in the inset of Fig. 6, is the low-temperature dependence of the magnetic entropy (per Ce atom) in $Ce_{11}Pd_4In_9$, obtained by integrating the $\frac{C_{4f}}{T}(T)$ data. The entropy released by $T_C$ equals 4.9 J/(mol$_{Ce}$ K) that is $0.85R\ln2$. The reduction of $S_{4f}$ with respect to the value expected for a doublet ground state can be attributed to the Kondo interactions. Applying the Bethe Ansatz solution (for the effective spin $S = ½$), in which the Kondo temperature $T_K$ is defined as the temperature where the entropy attains $0.65R\ln2$, one finds from Fig. 6 the value $T_K = 15$ K. Remarkably, the so-obtained magnitude of the Kondo temperature is close to that derived from Eq. 2.

*3.3. Electrical transport*

Fig. 7 displays the temperature variation of the electrical resistivity of $La_{11}Pd_4In_9$. The compound exhibits metallic conductivity with the resistivity of about 115 μΩ cm at room temperature and about 12 μΩ cm at liquid helium temperature. In the entire temperature range, except for $T < 0.6$ K, the experimental $\rho(T)$ curve can be approximated by the Bloch-Grüneisen-Mott (BGM) formula [28]

$$\rho(T) = \rho_0 + 4R\theta_R \left(\frac{T}{\theta_R}\right)^5 \int_0^{\frac{\theta_R}{T}} \frac{x^5 dx}{(e^x - 1)(1 - e^{-x})} + KT^3 \qquad (3)$$

where the first term is the residual resistivity due to scattering conduction electrons on static defects in the crystal lattice, the second term (hereafter labeled $\rho_{BGM}$) represents electron-phonon scattering processes ($\theta_R$ is sometimes considered as a rough measure of the Debye temperature), whereas the third term accounts for Mott-type inter band scattering. The BGM parameters obtained from least-squares fitting Eq. 3 to the experimental data of $La_{11}Pd_4In_9$ are: $\rho_0$ = 12.4(4) $\mu\Omega$cm, $\theta_R$ = 97.7(5) K, $R$ = 0.404(9) $\mu\Omega$cm/K and $K = -7.9(3) \times 10^{-7}$ $\mu\Omega$cm/K$^3$.

As shown in more detail in the inset to Fig. 7, the measured sample of $La_{11}Pd_4In_9$ was found superconducting below 0.6 K. In view of the lack of any corresponding anomaly in the heat capacity data (see above), one can rule out any supposition that superconductivity is an intrinsic bulk property of $La_{11}Pd_4In_9$. Probably, the effect appeared due to a tiny amount (not detected on the XRD pattern) of undefined foreign phase located at grain boundaries and/or surface of the specimen studied.

The temperature variation of the electrical resistivity of $Ce_{11}Pd_4In_9$ is shown in Fig. 8. The $\rho(T)$ curve is typical for Ce-based Kondo lattice systems. At room temperature, the resistivity equals 114 $\mu\Omega$cm, and it decreases with decreasing temperature down to about 2 $\mu\Omega$cm at $T$ = 0.38 K. The resultant residual resistivity ratio RRR = 57 is very large, which proves high quality of the polycrystalline sample studied. A clear bending in $\rho(T)$, observed near 150 K, is an indicative of the strong CEF interactions, in concert with the specific heat data (see above). The ferromagnetic phase transition manifests itself as a sudden drop in the resistivity below $T_C$ = 18.6 K, caused by reduction in spin-disorder scattering. At lower temperatures, $\rho(T)$ forms a faint maximum at around 5 K that is followed by another sharp decrease in its magnitude. These features can be attributed to a crossover from incoherent to coherent Kondo regime, and the order-order transition at $T_t$ = 1.7 K. The two magnetic singularities are clearly evident in the low-temperature variation of the temperature derivative of the resistivity, presented in the inset to Fig. 8. It should be noted that the overall behavior of $d\rho/dT(T)$ matches perfectly well with the $C(T)$ data of $Ce_{11}Pd_4In_9$ (compare Fig. 5), which revealed the second-order phase transition at $T_C$ and the first order transition at $T_t$.

Assuming that $Ce_{11}Pd_4In_9$ and $La_{11}Pd_4In_9$ have similar phonon spectra, the non-phonon contribution to the electrical resistivity of the former compound was found by subtracting from its measured $\rho(T)$ data the $\rho_{BGM}(T)$ component derived for the latter material from Eq. 3. The so-estimated sum of the magnetic component due to 4$f$ electrons, $\rho_{4f}(T)$, and the residual resistivity, $\rho_0$, in $Ce_{11}Pd_4In_9$ is displayed in Fig. 9. In the paramagnetic region, the behavior of $\rho_{4f}(T)$ is clearly governed by an interplay of the Kondo and CEF effects, as expected for Kondo lattices [29]. Above about 150 K, the experimental data can be approximated by the formula

$$\rho_{4f}(T) + \rho_0 = (\rho_0 + \rho_0^\infty) + c_K \ln T \quad , \qquad (4)$$

where $\rho_0^\infty$ stands for the spin-disorder resistivity and $c_K$ is the Kondo coefficient. The least-squares fitting yielded the parameters: $(\rho_0 + \rho_0^\infty) = 113.1(7)$ μΩcm and $c_K = -17.4(2)$ μΩcm. Setting $\rho_0 \approx 2$ μΩcm, as measured at $T = 0.38$ K, one can estimate the magnitude of $\rho_0^\infty$ to be about 100 μΩcm. This value should be considered as the upper limit for the spin-disorder resistivity, derived without taking into account the CEF interactions. In turn, the fairly large magnitude of $c_K$ corroborates the significance of the Kondo effect in $Ce_{11}Pd_4In_9$.

Below $T_C$, yet in the incoherent Kondo state (i.e. above 5 K), the $\rho_{4f}(T)$ data were analyzed in terms of scattering conduction electrons on spin-waves excitations. Applying the formula [22,30,31]

$$\rho_{4f}(T) + \rho_0 = (\rho_0 + \rho_m) + BT\Delta\left(1 + \tfrac{2T}{\Delta} + \tfrac{1}{2}e^{\tfrac{-\Delta}{T}}\right)e^{\tfrac{-\Delta}{T}} \quad , \qquad (5)$$

where $\rho_m$ denotes the incoherent magnetic scattering contribution, and the second term represents the electron-magnon scattering processes with $\Delta$ being an energy gap in the spin-waves spectrum, one obtained a good description of the resistivity data of $Ce_{11}Pd_4In_9$ (see Fig. 9) with the parameters: $\rho_m = 10.2(1)$ μΩcm ($\rho_0 = 2$ μΩcm was assumed), $B = 0.097(2)$ μΩcm/K$^2$ and $\Delta = 51.6(5)$ K. It is worth noting that the value of $\rho_m$ is much smaller than $\rho_0^\infty$, as anticipated for residual CEF contribution to the resistivity at low temperatures.

As can be inferred from Fig. 9, below $T_t$, the electrical resistivity of $Ce_{11}Pd_4In_9$ is dominated by electron-electron scattering processes leading to the Fermi liquid type behavior $\rho_{4f}(T) \sim T^2$. This finding corroborates a strongly correlated electronic ground state in the compound investigated.

## 4. Conclusions

The novel compound $Ce_{11}Pd_4In_9$ crystallizes in the orthorhombic $Nd_{11}Pd_4In_9$-type structure (space group *Cmmm*), with as many as five inequivalent sublattices for the Ce atoms. It orders ferromagnetically at $T_C = 18.6$ K and undergoes another magnetic phase transition at $T_t = 1.7$ K. In the entire temperature range studied, the thermodynamic and electrical transport properties of this material are governed by an interplay between Kondo and CEF interactions.

At the lowest temperatures, the behavior in $Ce_{11}Pd_4In_9$ is determined by strong electron-electron interactions arising in the ferromagnetic state.

**Acknowledgment**

The work was supported by the National Science Centre (Poland) under research grant No. 2015/19/B/ST3/03158.

**References**

[1] G. R. Stewart; Rev. Mod. Phys. **56**, 755 (1984).

[2] G. R. Stewart; Rev. Mod. Phys. **73**, 797 (2001).

[3] F. Steglich; J. Phys. Soc. Jpn **74**, 167 (2005).

[4] D. Kaczorowski, A. P. Pikul, D. Gnida, and V. H. Tran; Phys. Rev. Lett. **103**, 27003 (2009).

[5] P. Gegenwart, Q. Si and F. Steglich; Nature Phys. **4**, 186 (2008).

[6] C. Pfleiderer; *Rev. Mod. Phys.* **81**, 1551(2009).

[7] ] C. Krellner, N.S. Kini, E.M. Bruning, K. Koch, H. Rosner, M. Nicklas, M. Baenitz, C. Geibel; Phys. Rev. B **76** ,104418 (2007).

[8] M. Matusiak, A. Lipatov, A. Gribanov, D. Kaczorowski; J. Phys. Condens. Matter **25**, 265601 (2013).

[9] A. Tursina, E. Khamitcaeva, D. Gnida, D. Kaczorowski; J. Alloys Compd. **731**, 229 (2018).

[10] D. Das, T Gruner, H Pfau, U B Paramanik, U Burkhardt, C Geibel and Z Hossain; J. Phys.: Condens. Matter **26**, 106001 (2014).

[11] D. Das, S. Nandi, I. da Silva, D. T. Adroja and Z. Hossain, Phys. Rev. B **94**, 174415 (2016).

[12] J. Tang, K. A. Gschneidner Jr., Steven J. White, M. R. Roser, T.J. Goodwin, and L. R. Corruccini; Phys. Rev. B **52**, 7328 (1995).

[13] J. Custers , K-A. Lorenzer, M. Müller,  A. Prokofiev, A. Sidorenko, H. Winkler, A. M. Strydom, Y. Shimura, T. Sakakibara, R. Yu, Q. Si and S. Paschen; Nature Mater. **11**, 189 (2012).

[14] A. Strydom, A. Pikul, F. Steglich and S. Paschen; J. Phys.: Conf. Ser. **51**, 239 (2006)

[15] J. Prokleška, M. Kratochvílová, K. Uhlířová, V. Sechovský and J. Custers;  Phys. Rev. B, **92**, 161114 (2015).

[16] L. Sojka, M. Manyako, R. Černý, M. Ivanyk, B. Belan, R. Gladyshevskii a , Ya Kalychak; Intermetallics **16,** 625 (2008).

[17] M. Pustovoychenko, Yuriy Tyvanchuk, Ivan Hayduk, Yaroslav Kalychak; Intermetallics **18**, 929 (2010).

[18] L. Sojka, M. Demchyna, B. Belan, M. Manyako and Ya. Kalychak; Intermetallics  **49**, 14 (2014).

[19] A. Szytuła, S. Baran, J. Przewoźnik, Yu. Tyvanchuk and Ya. Kalychak; J. Alloys Compd. **601,** 238 (2014).

[20] A. Szytuła, S. Baran, B. Penc, J. Przewoźnik, A. Winiarski, Yu. Tyvanchuk and Ya.M. Kalychak; J. Alloys Compd. **589**, 622 (2014).

[21] V. Gribanova, E. Murashova, D. Gnida, Zh. Kurenbaeva, S. Nesterenko, A. Tursina, D. Kaczorowski and A. Gribanov; J. Alloys Compd. **711**, 455 (2017).

[22] A. Prasad, V. K. Anand, U. B. Paramanik, Z. Hossain, R. Sarkar, N. Oeschler, M. Baenitz, and C. Geibel; Phys. Rev. B, **86**, 014414 (2012)

[23] P. Manfrinetti, S. K. Dhar, R. Kulkarni and A. V. Morozkin;  Solid State Commun.**135**, 444 (2005).

[24] E.S.R. Gopal;  Specific Heats At Low Temperatures (Plenum Press, New York, 1966).

[25] M. J. Besnus, A. Braghta, N. Hamdaoui and A. Meyer; J. Magn. Magn. Mater. **104-107**, 1385 (1992).

[26]  A. Blanco, M. de Podesta, J. I. Espeso, J. C. Gómez Sal, C. Lester, K. A. McEwen, N. Patrikios, J. Rodríguez Fernández; Phys. Rev. B **49**, 15126 (1994).

[27] M. Szlawska and D. Kaczorowski; Phys Rev B **85**, 134423 (2012).

[28] N. F. Mott and H. Jones; The Theory of the Properties of Metals and Alloys(Oxford University Press, London, 1958).

[29] B. Cornut and B. Coqblin; Phys. Rev. B **5**, 4541 (1972).

[30] N. H. Andersen and H. Smith; Phys. Rev. B **19**, 384 (1979).

[31] N. H. Andersen, P. E. Gregers-Hansen, E. Holm, H. Smith and O. Vogt; Phys. Rev. Lett. **32**, 1321 (1974).

**Figure captions**

Fig. 1. Temperature dependence of the inverse molar magnetic susceptibility of $Ce_{11}Pd_4In_9$ measured in an applied magnetic field of 1 kOe. The solid red line represents the Curie Weiss fit discussed in the text. Inset: low-temperature variation of the magnetization in $Ce_{11}Pd_4In_9$ taken in a field of 1 kOe, upon cooling the sample in the applied field.

Fig. 2. Magnetic field variation of the magnetization in $Ce_{11}Pd_4In_9$ measured at 1.7 K with increasing (full circles) and decreasing (open circles) field strength. Inset: (in a magnified scale). Inset: zoom-in of the isothermal magnetization data up to 4 kOe.

Fig. 3. Temperature dependencies of the magnetization in $Ce_{11}Pd_4In_9$ measured in an applied magnetic field of 20 Oe upon cooling the sample in zero field (ZFC) and applied field (FC).

Fig. 4. Temperature dependence of the specific heat of $La_{11}Pd_4In_9$. The solid red curve represents the fit discussed in the text. Inset: low-temperature data in the form of specific heat over temperature versus squared temperature. The solid straight line is the Debye fit discussed in the text.

Fig. 5. Temperature dependence of the specific heat of $Ce_{11}Pd_4In_9$. The solid line represents the phonon contribution discussed in the text. Inset: low-temperature data in the form of specific heat over temperature versus temperature.

Fig. 6. Temperature variation of the 4*f*-derived specific heat of $Ce_{11}Pd_4In_9$ (note a semi-logarithmic scale). Inset: temperature variation of the magnetic entropy in $Ce_{11}Pd_4In_9$.

Fig. 7. Temperature dependence of the electrical resistivity of $La_{11}Pd_4In_9$. The solid red curve represents the Bloch-Grüneisen-Mott fit discussed in the text. Inset: low-temperature resistivity data.

Fig. 8. Temperature dependence of the electrical resistivity of $Ce_{11}Pd_4In_9$. Inset: low-temperature dependence of the temperature derivative of the resistivity data.

Fig. 9. Temperature dependence of the non-phonon electrical resistivity of $Ce_{11}Pd_4In_9$ (note a semi-logarithmic scale). The solid red line and long-dashed pink lines represents the fits of Eq. 4 and Eq. 5 to the experimental data, respectively, as discussed in the text. The short-dash dark-red curve marks a Fermi liquid type behavior.

Fig. 1

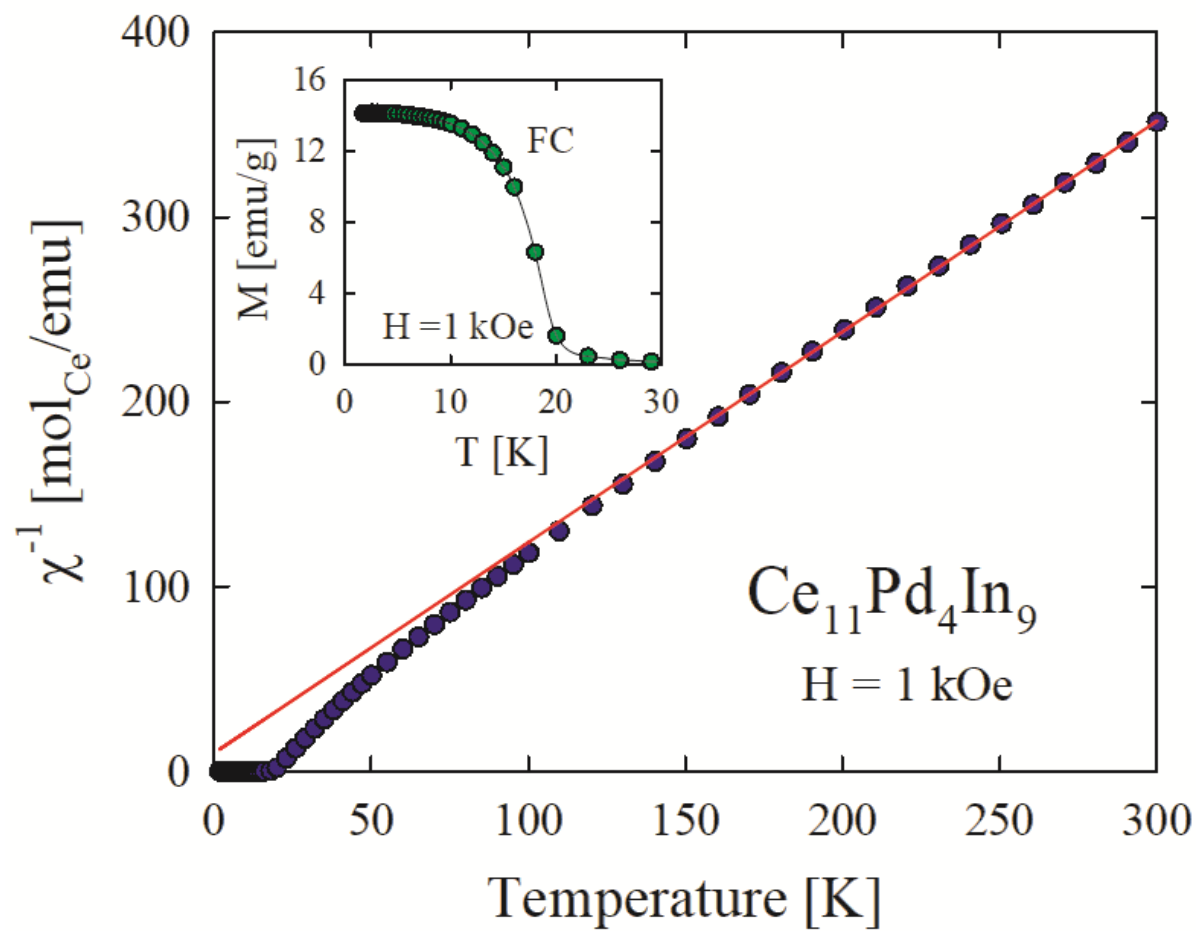

Fig. 2

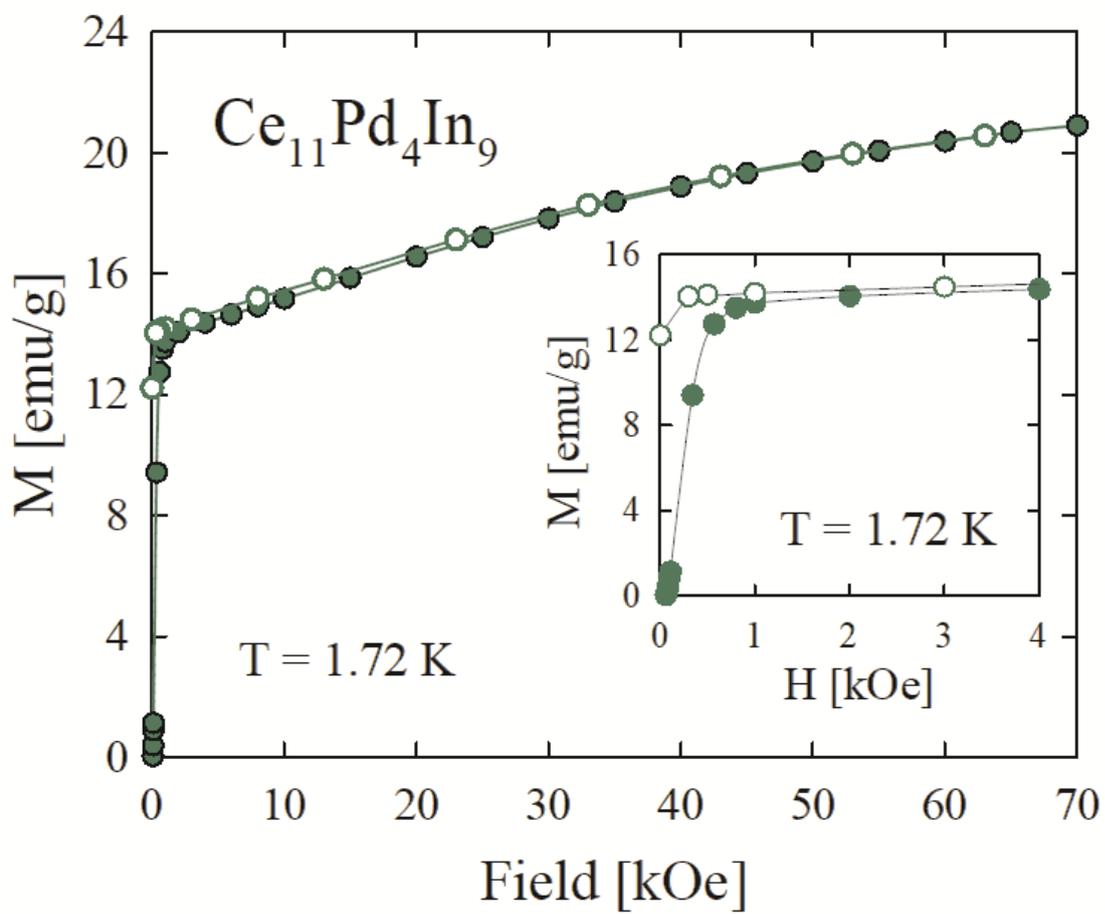

Fig. 3

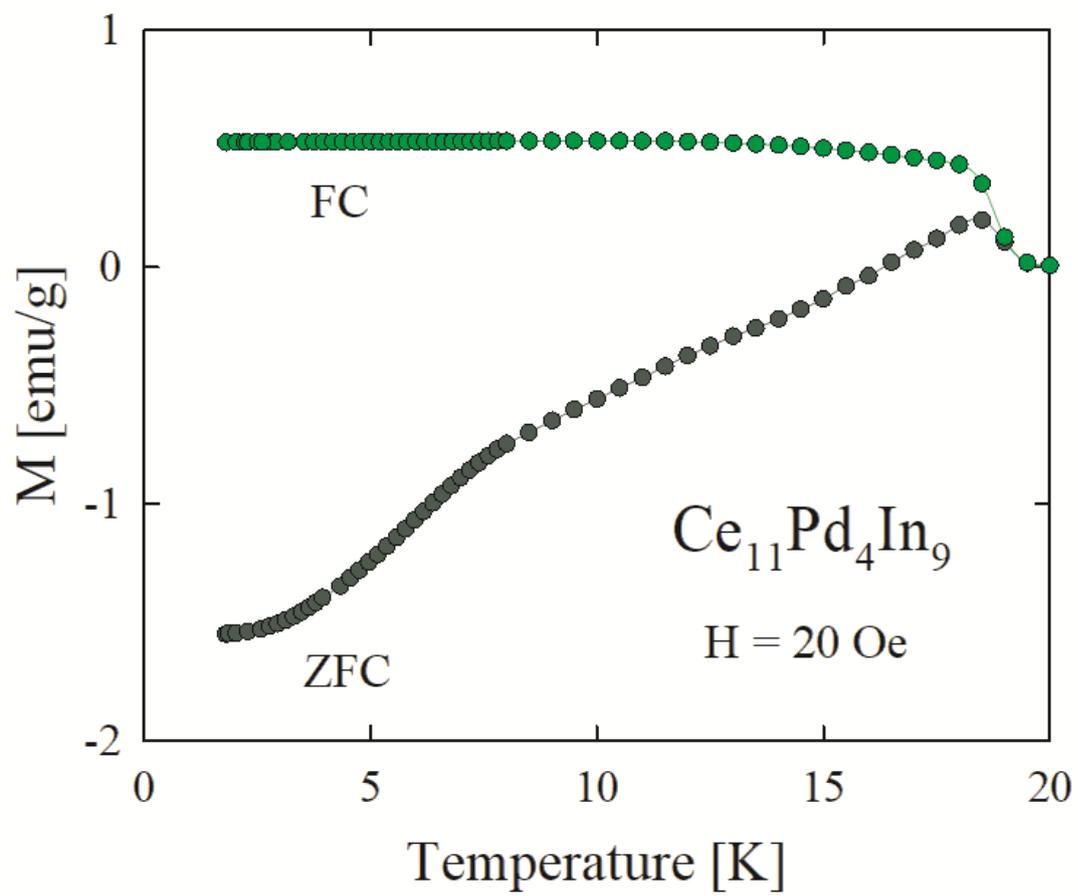

Fig. 4

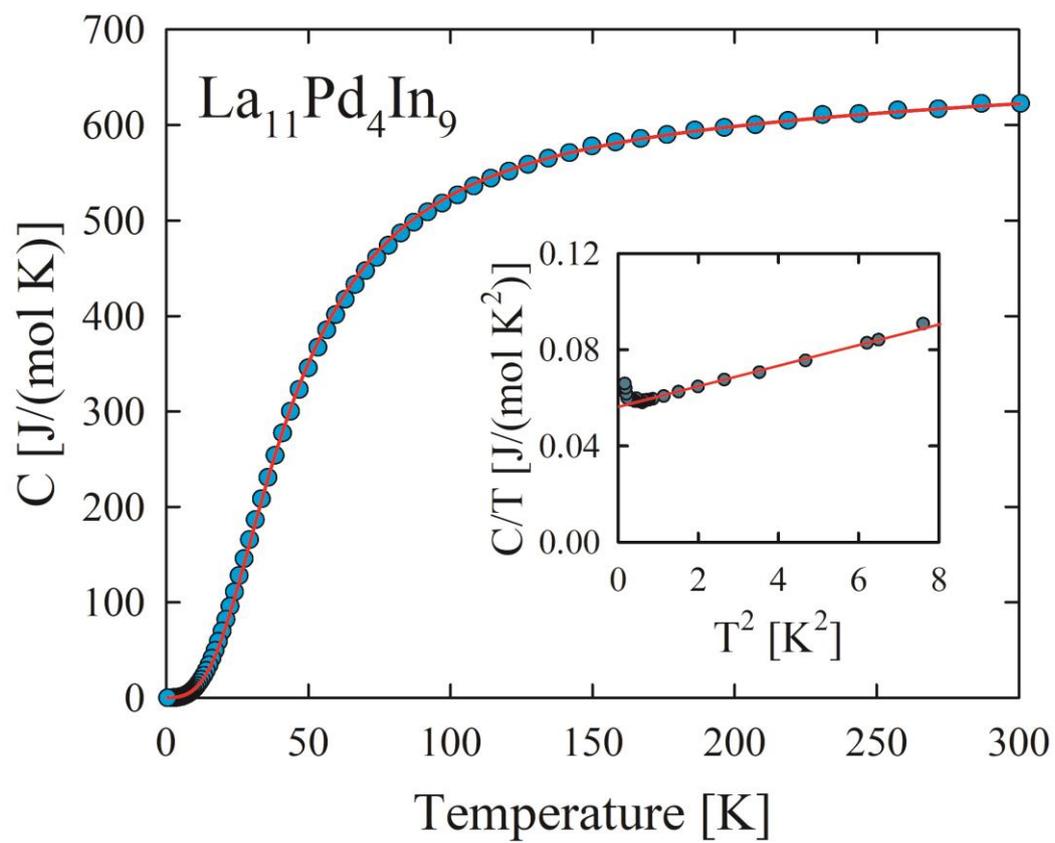

Fig. 5

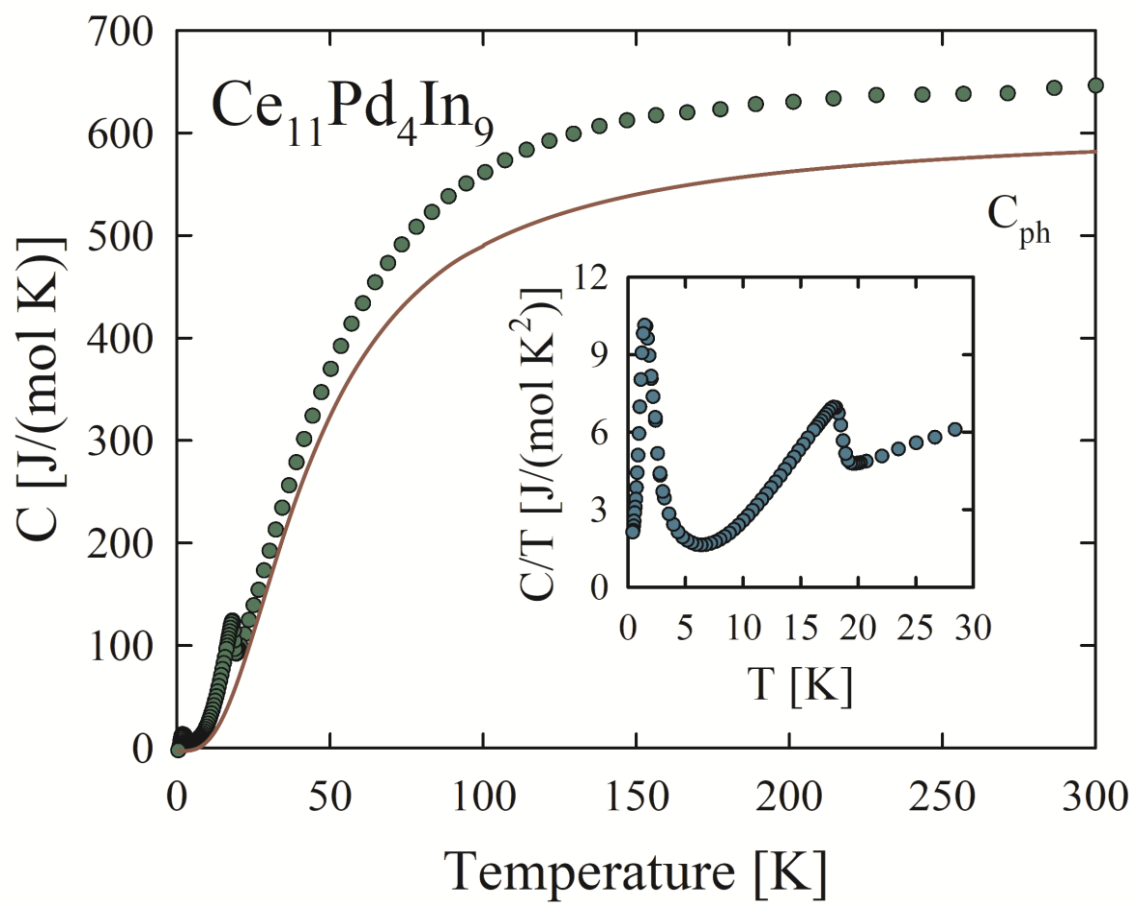

Fig. 6

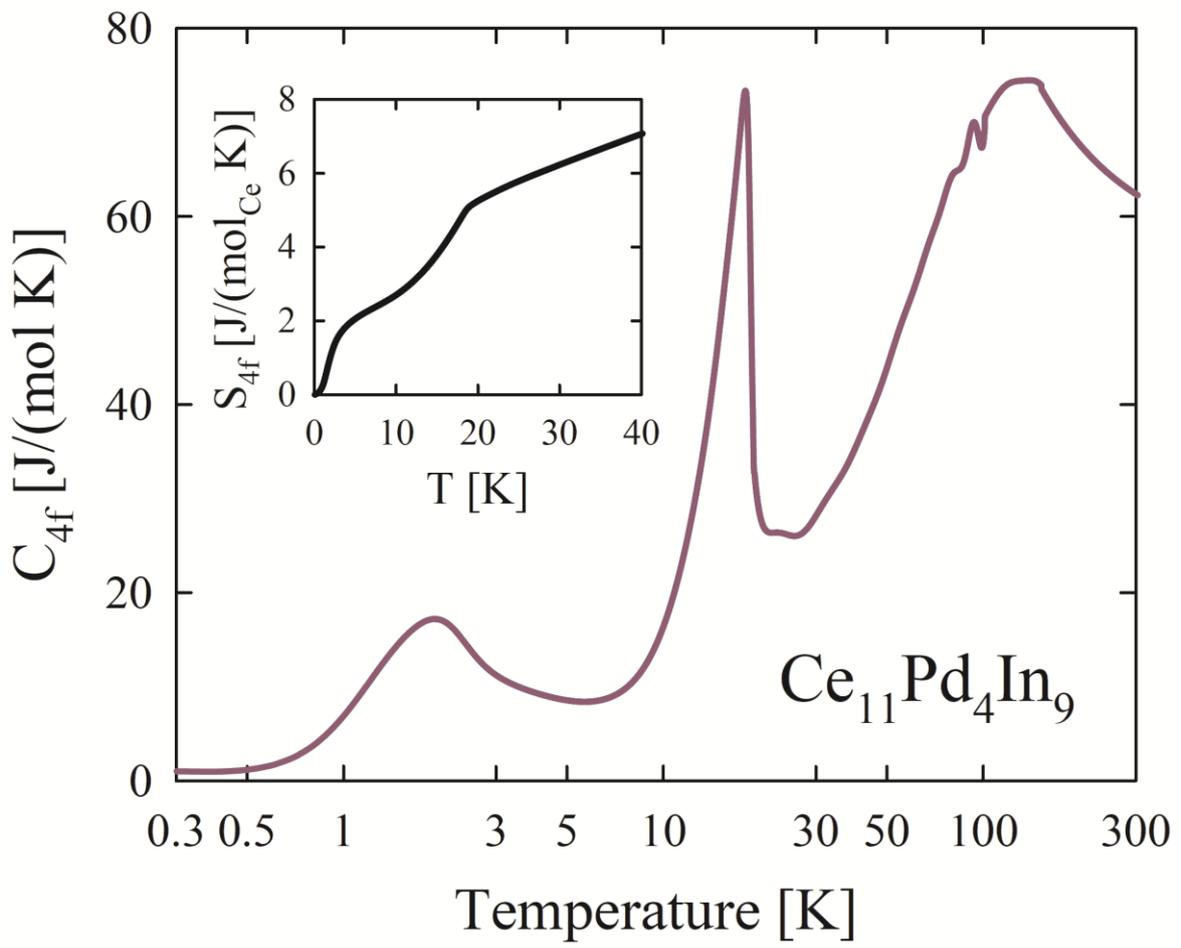

Fig. 7

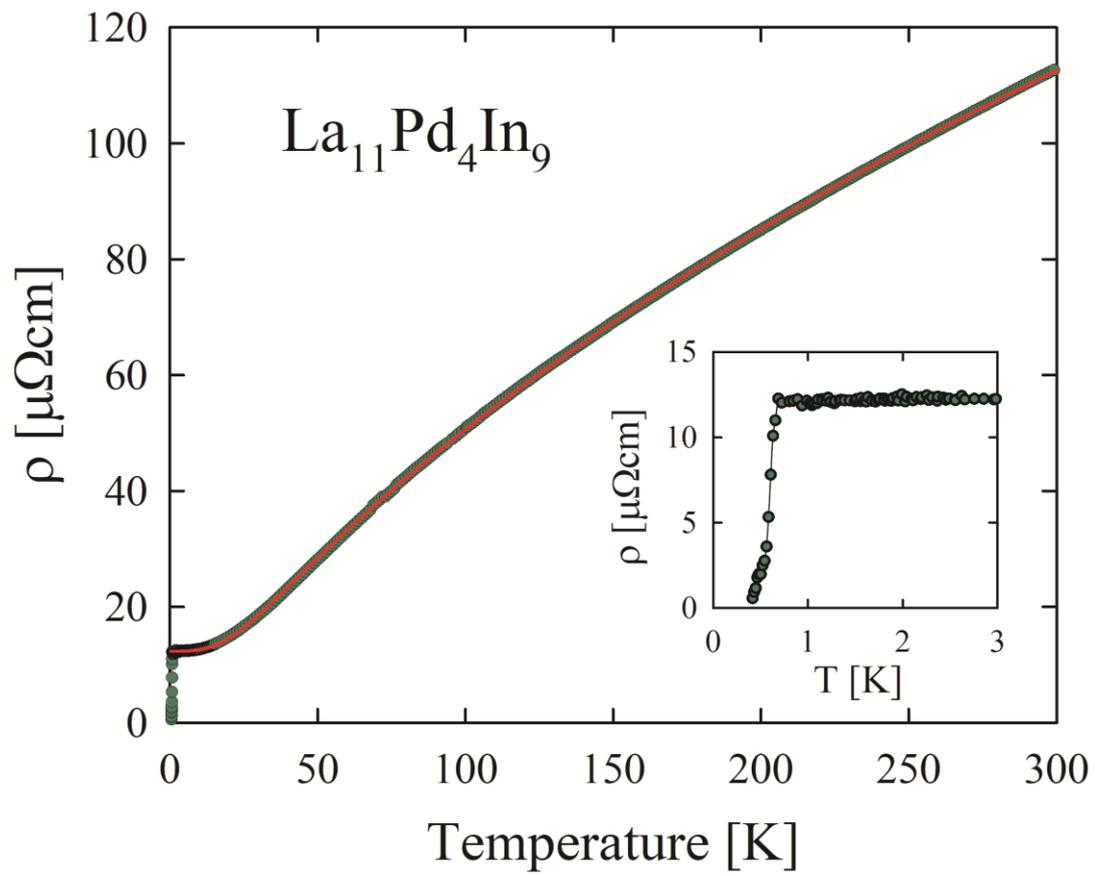

Fig. 8

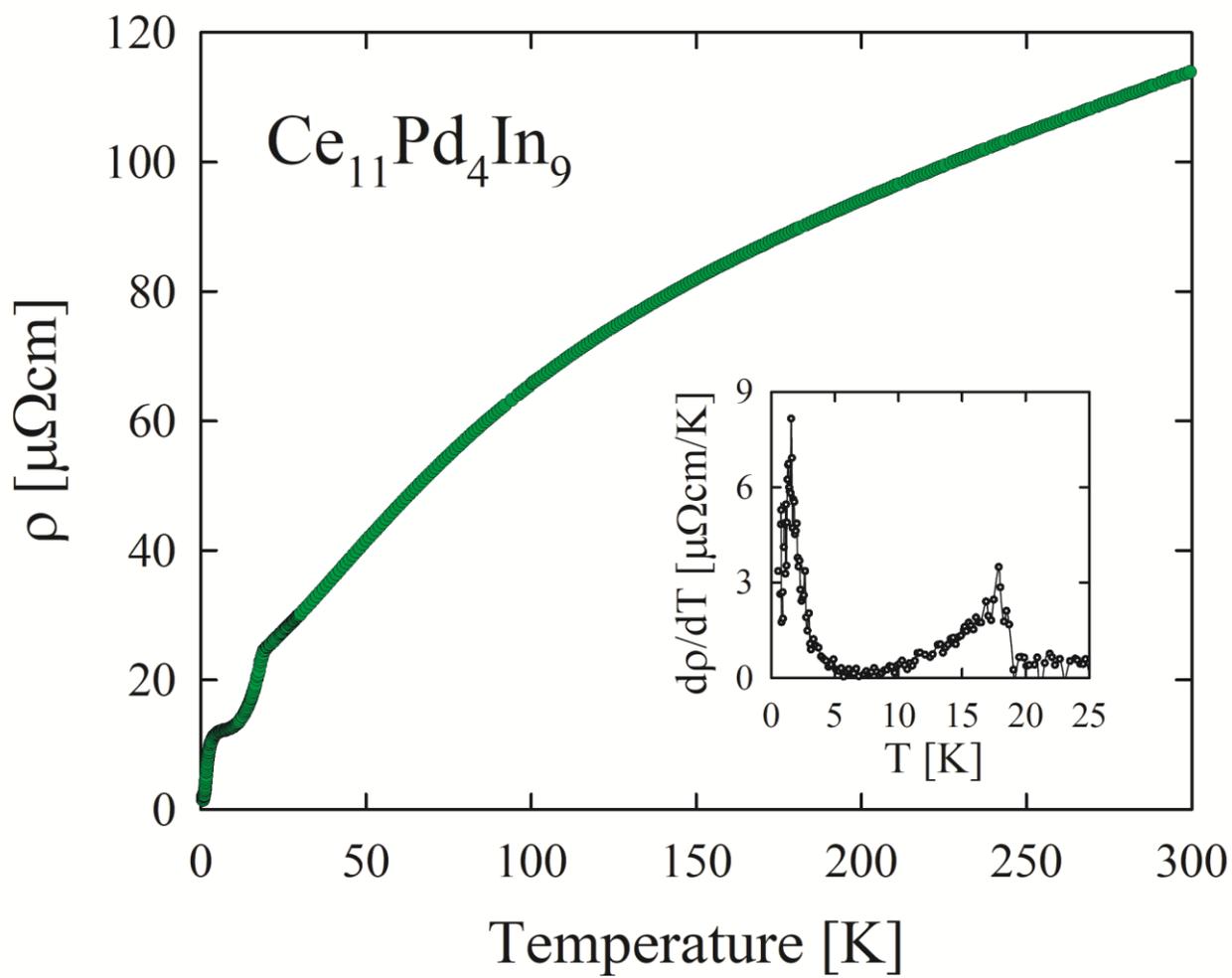

Fig. 9

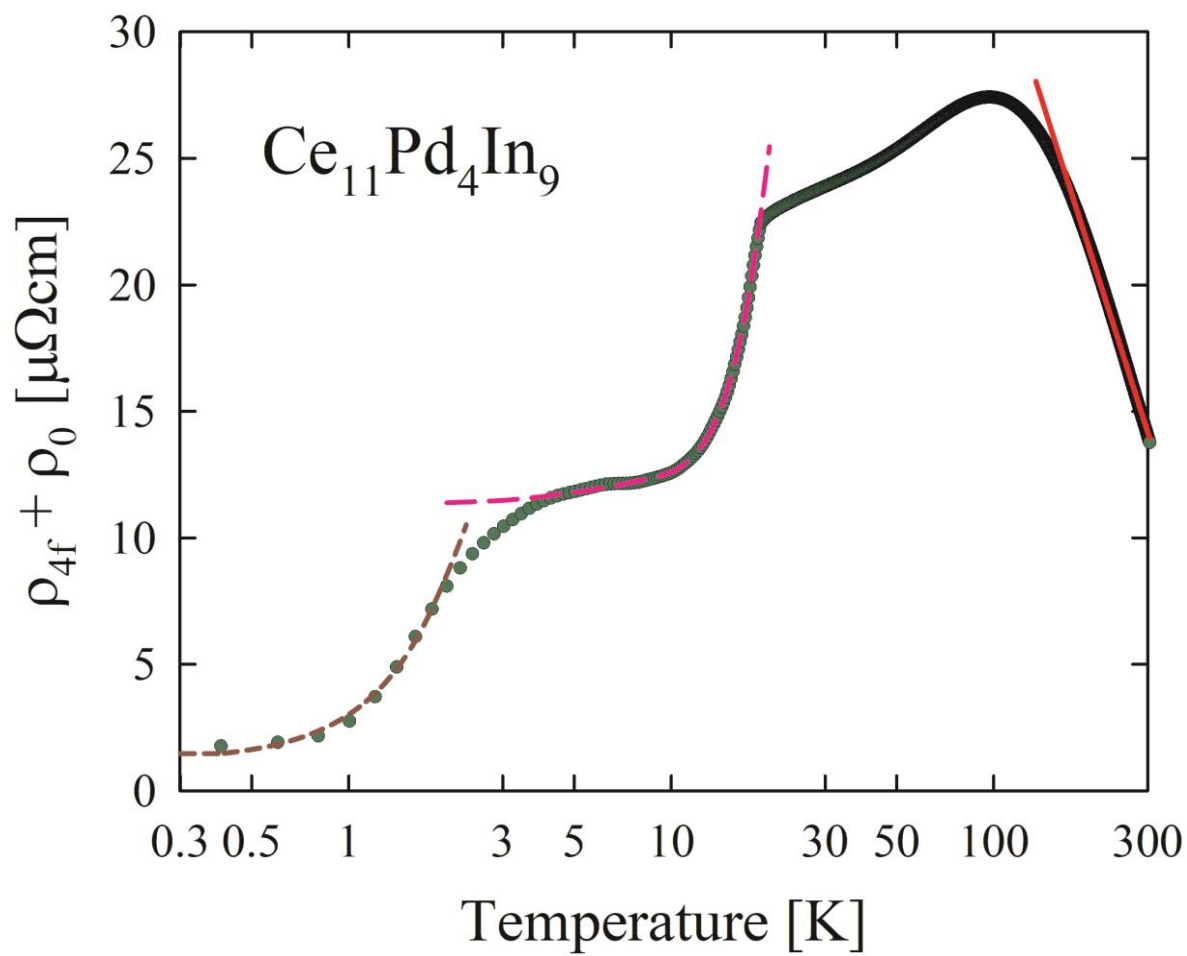